# Visual Exploratory Data Analysis of the COVID-19 Vaccination Progress in Nigeria

Orji, .U. E., Ugwuishiwu, .C. H., Okoronkwo, M.C., Asogwa, .C.N., Ogbene, .N.

**Abstract**— The coronavirus outbreak in 2020 devastated the world's economy, including Nigeria, even resulted in a severe recession. Slowly the country is building back again, and the vaccines are helping to reduce the spread of COVID-19. Since the COVID-19 vaccine came to Nigeria in March, 2021; 18,728,188 people have been fully vaccinated as at May 31st, 2022. This is roughly 10% of the Nigerian population estimated at 206.7 million [1]. This paper presents a visual Exploratory Data Analysis of the COVID-19 vaccination progress in Nigeria using the R-tidyverse package in R studio IDE for data cleaning & analysis, and Tableau for the visualizations. Our dataset is from the Nigerian National Primary Health Care Development Agency (NPHCDA) in charge of the vaccines. The data used for this research contain the state-by-state breakdown of COVID-19 vaccine distribution recorded between March 5th, 2021, and May 31st, 2022. This paper aims to show how these data analytics tools and techniques can be useful in finding insights in raw data by presenting the results of the EDA visually thus reducing the ambiguity and possible confusions that is associated with data in tables. Furthermore, our findings contribute to the growing literature on COVID-19 research by showcasing the COVID-19 vaccination trend in Nigeria and the state by state distribution.

**Index Terms**— Data Analytics, EDA, Visual EDA, R-tidyverse, Tableau Dashboards, COVID-19 Vaccine.

—————— ◆ ——————

## 1 INTRODUCTION

According to a recent Statista report [2], Nigeria has the highest population in Africa (estimated to be over 200 million). It is no wonder Nigeria is one of the epicenters of COVID-19 in Africa. The first case of COVID-19 in Nigeria was reported on February 27th, 2020. According to the Nigeria Centre for Disease Control, it was also the very first documented case in sub-Saharan Africa [3].

The country has fought hard to control the outbreak and mitigate its impact on all sectors of the economy. As at May 27th, 2022, they have been a total of 256,028 confirmed cases, and 250,036 recoveries but; unfortunately, 3,143 deaths have been recorded. There are 2,940 active cases as at May 27th, 2022 [4].

According to [5], as of November 25th, 2021; 223 countries have authorized the use of 22 COVID-19 vaccines from different manufacturers (4 authorized and 18 with emergency use permission) to vaccinate their citizens.

The COVID-19 vaccination in Nigeria started on March 5th, 2021 [6]. According to the latest data from the National Primary Health Care Development Agency (NPHCDA) for May 31st, 2022, they have recorded 18,728,188 fully (1st & 2nd dose) vaccinated people, and 26,887,802 people has taken only the first dose [7].

The World Health Organization (WHO) Strategic Advisory Group of Experts (SAGE) on Immunization proposed a 3-stage prioritization roadmap for vaccine administration with the aim of guiding countries with the COVID-19 vaccine distribution as a result of the limitation in vaccine supply. The 3-stage prioritization roadmap include;

- Stage 1 = 1-10 % of a country's total population is to be vaccinated.
- Stage 2 = 11-20% is to be vaccinated and;
- Finally, at Stage 3 = 21-50% of a country's population is to be vaccinated [8].

Ideally, when handling a data analysis problem, regardless of how simple or complex the problem is, the first step should be to examine the data properly; this process is known as Exploratory Data Analysis (EDA). According to [9], the primary purpose of any EDA is to get familiar with the data. During EDA, you are likely to detect problems in the data, including outliers, non-normal distributions, missing values, or other errors, which, if left unhandled, will skew the result and damage the integrity of your findings [10] [11].

The research by [12] is credited for coining the term "Exploratory Data Analysis." The EDA process is a crucial component of any Data Mining activity per the industry standard Cross Industry Standard Process for Data Mining (CRSP-DM) framework. CRSP-DM was developed to help data analysts plan, organize, and implement their data analysis projects [13].

EDA is categorized into two techniques; the descriptive statistics and graphical techniques [14]. The descriptive statistics technique includes the various data manipulation tasks needed to shape the data and ensure it is fit for purpose, whereas the graphical technique comprises the different visualizations. Both of these techniques ensure that the data is understood by showing the trends, patterns, and relationships between the variables.

This paper looks at the COVID-19 vaccination data in Nigeria and performs Exploratory Data Analysis to show

————————————————
**Author's Details**
- *Orji, .U. E. is a post-graduate student of Department of Computer Science, University of Nigeria, Nsukka. E-mail: ugochukwu.orji.pg00609@unn.edu.ng*
- *Ugwuishiwu, .C. H. is with the Department of Computer Science, University of Nigeria, Nsukka. E-mail: chikodili.ugwuishiwu@unn.edu.ng*
- *Okoronkwo, M.C. is with the Department of Computer Science, University of Nigeria, Nsukka. E-mail: Matthew.okoronkwo@unn.edu.ng*
- *Asogwa, .C.N. is with the Department of Computer Science, University of Nigeria, Nsukka. E-mail: caroline.asogwa@unn.edu.ng*
- *Ogbene, .N. is with the Department of Computer Science, University of Nigeria, Nsukka. E-mail: nnaemeka.Ogbene@unn.edu.ng*

the progress report, trends and impact. We used tidyverse package of R programming to perform EDA on the data and Tableau to create the visualizations/dashboard.

## 2 LITERATURE REVIEW

The COVID-19 vaccination process in Nigeria has been plagued with different but usual challenges like infrastructural inadequacies, primitive technological capacity, and lack of skilled workforce in the Nigerian cold chain system. These challenges are made even more difficult due to the storage peculiarities of the Pfizer and Moderna COVID-19 vaccines. In [15], the authors analyzed the current status of the Nigerian cold chain system with emphasis on workers' technical capacity in the various geopolitical zones in Nigeria. The authors highlighted the challenges faced by Nigeria in vaccine procurement and also proffered possible solutions to the defective cold-chain system and ways to improve the efficient distribution of the COVID-19 vaccines in Nigeria.

There have a recent boom in automation of the data analysis process over the past few years. This is especially important because of the increasing availability of heterogeneous and, in most cases, opaque datasets from varied sources. EDA is crucial to help analysts understand the data, perform cleaning and validation checks, and carry out feature engineering where necessary. The authors in [16] presented a systematic review of currently available tools for performing automated EDA (autoEDA) in R. The research explored important features in fifteen (15) R packages that effectively help automate the EDA process.

Furthermore, one of the key advantages of EDA is that it presents analysts with a tremendous opportunity to identify undiscovered knowledge from raw data. Data analysis tools like R makes it even easier for non-programmers to get insights from massive datasets through interactive visualizations like line charts, box plots, histograms etc. Researchers are also creating their own data analysis tools. For example, [17] developed a MetaOmGraph (MOG), open-source, standalone software for performing EDA on massive datasets. The authors successfully demonstrated the efficiency of their MOG on a case study of large human cancer RNA-Seq dataset to identify novel putative biomarker genes in different tumors.

Also, Informatics tools that help improve processes and productivity are a growing demand for bioinformatics researchers whose data comprise soil, plant, and aquatic samples, which usually contain complex mixtures of proteins, carbohydrates, lipids, hydrocarbons, and other compounds. Techniques like the Fourier transform mass spectrometry (FT-MS) is making it possible for these researchers to get a high-resolution and accurate reading from such data. A paper by [18] presented ftmsRanalysis, an R package for analyzing FT-MS data. The authors used a case study of soil microbiology to demonstrate the core functionality and highlighted the capabilities of their ftmsRanalysis package.

To build a machine learning or statistical model, you must first perform EDA as an initial investigation process to know more about your data via descriptive statistics and visualizations. However, EDA can be a very tedious task, thus the need to automate the EDA process. The authors in [19] proposed a SmartEDA for R that automates the EDA process. The authors also compared their SmartEDA with other packages available for EDA in the Comprehensive R Archive Network (CRAN).

Another great tool for EDA is Tableau, which is an extremely powerful data visualization tool capable of handling the analysis and visualization of massive datasets. It has a drag and drop interface that makes it easy to use and produce great visualizations and dashboards. The platform also supports varieties of data sources and structures. In [20], the authors effectively deployed Tableau to present a comprehensive visual EDA of the global COVID-19 data. Furthermore, Data visualization over the past few years has become increasingly vital in telling the data story as it intuitively shows the results of data analysis. Modern data visualization tools like Tableau also help in the whole process of collecting, cleaning, analyzing, and sharing data. Most significantly, in this era of COVID-19 and the availability of vast amounts of data, visualization helps people understand the data. A similar paper by [21] demonstrated this by using Tableau to analyze India's COVID-19 healthcare data and created highly impressive visualizations to help tell the story. Finally, the vaccines are obviously the next big step in combating the COVID-19 pandemic. Also, the vaccination campaign is gaining widespread acceptance as more people globally are accessing and taking the vaccine. A related paper by authors in [5] carried out an EDA of the global vaccination distribution using various python libraries for analysis and visualization. The authors showed a link between increasing vaccinations and lowering COVID-19 spread around the world.

### 2.1 Research gap

After reviewing the related works done in this area, we established that the use of data analytics techniques and tools for EDA is on the rise and the results achieved has been promising. However, there have also been limited research on Nigeria's COVID-19 vaccination status. Therefore, in this paper, we deployed two highly effective tools, R-tidyverse and Tableau for EDA and focused on the vaccination progress in Nigeria as reported between March 5th, 2021, and May 31st, 2022.

Table 1: Comparison of similar studies

| Author | Title | Concept handled | This paper |
|---|---|---|---|
| [5] | Global landscape of COVID-19 vaccination progress: insight from an exploratory data analysis | EDA of the global vaccination distribution using various python libraries for analysis and visualization | EDA of COVID-19 vaccine in Nigeria with R-tidyverse and Tableau. |
| [21] | Data Visualization View with Tableau. | EDA of Indian COVID-19 data with Tableau | |
| [20] | Data analytics and visualization using Tableau utilitarian for COVID-19 (Coronavirus). | Visual EDA of the global COVID-19 data with Tableau | |
| [19] | SmartEDA: An R package for automated exploratory data analysis. | Proposed a SmartEDA package for automating the EDA process in R | |
| [18] | ftmsRanalysis: An R package for exploratory data analysis and interactive visualization of FT-MS data. | Developed ftmsRanalysis R package for analysing FT-MS data. | |
| [17] | MetaOmGraph: a workbench for interactive exploratory data analysis of large expression datasets. | Developed a MetaOmGraph (MOG) an open-source, standalone software for performing EDA on massive datasets. | |
| [16] | The landscape of R packages for automated exploratory data analysis. | Systematic review of currently available tools for performing automated EDA (autoEDA) in R. | |

## 3 METHODOLOGY

This section describes our dataset, the techniques for preprocessing and cleaning the data, and Tableau visualization software used to tell the data story.

### 3.1 Dataset Description
In this research, we used the Nigerian National Primary Health Care Development Agency (NPHCDA) dataset, which is publically accessible online [22]. The dataset comprises records of COVID-19 vaccinations in all states of Nigeria between March 5th, 2021, and May 31st, 2022. The dataset consists of 5 columns (State, Latitude, Longitude, Population, Total Vaccinated Population, Population vaccinated (1st Dose), Population vaccinated (2nd Dose)) and 38 rows with the data from all 36 states plus the FCT. Table 2 gives a brief description of the dataset.

Table 2: Brief description of the dataset

| Variable | Description | Data Types |
|---|---|---|
| State | All states in Nigeria plus FCT | string |
| Latitude | Geographical coordinates of states | float64 |
| Longitude | Geographical coordinates of states | float64 |
| Population | Total population of each state | int64 |
| Total Vaccinated Population | Total vaccinated population for each state | int64 |
| Population vaccinated (1st Dose) | Population that has taken the 1st dose for each state | int64 |
| Population vaccinated (2nd Dose) | Population that has taken the 2nd dose for each state | int64 |

### 3.2 Data Preprocessing and Cleaning using R-tidyverse package

The R programming language has an interesting set of very efficient packages for cleaning and preprocessing large datasets. The tidyverse package contains some of R programming's most useful functions for data tidying. Cleaning or tidying your dataset is an essential step you must take before analysis to ensure your data are in the format you need.

Most data analysis research projects will usually require some level of data manipulation; in most cases, the dataset we get online or from various sources are not clean and might not suit our purpose. Fortunately, R offers plenty of benefits via tidyverse including;

- Provides great data manipulation flexibility [23].
- R uniquely offers reproducibility of data analysis [24]
- It also provides fast and transparent bug fixing options
- R offers an impressive community of users in all fields of life which is especially important for beginners.

Fig 1 below shows the different stages of data cleaning and preprocessing in R tidyverse as the data goes from dirty and noisy raw data to clean and action-oriented information.

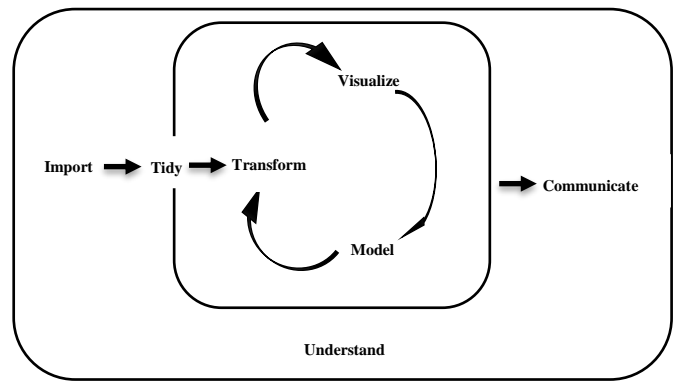

Fig 1: Stages of data preprocessing in R tidyverse [25]

An essential function of the tidyverse is to tidy data [26]. In R, every column is a variable, every row is an observation, and each cell contains a single value. However, as we already know that most datasets come in an unstructured and noisy state, the tidyverse core component "tidyr" package provides users with the tools to tidy it up and ensure the data is ready for analysis [27].

The data cleaning and preprocessing steps in this project include:

- Checking data types,
- Formatting classes,
- Checking for inconsistencies and missing values,
- Data mutation and aggregation.

### 3.3 Visualization with Tableau

Tableau is a very efficient data analysis and visualization tool capable of connecting with a variety of data sources. One of the most significant advantages of Tableau is that it offers users so many options to manipulate data. The Tableau software is straightforward to use, especially for non-programmers, as it enables the creation of dashboards without any coding knowledge through a visual, intuitive drag and drop interface. Today, organizations of all sizes utilize Tableau to translate their data into insightful visual dashboards [28]. With Tableau, anyone can create different visualizations to present their data and showcase insights interactively. Tableau's tools allow analysts to perform visual exploratory analysis (EVA) and see the data's visual impacts, which is easily understood. The Tableau software also has real-time data analytics capacity with 24/7 cloud support; it helps analysts enhance the data power [29]. Tableau is one of the few business intelligence tools designed for individual use but has since been scaled for businesses. There are varieties of visualizations created in Tableau according to user experience and needs [30].

## 4 RESULTS AND DISCUSSION

The results of our study from the EDA are presented in visualizations as shown below. The visualizations and dashboard were created on Tableau, the platform helps present EDA as visualization making it easier to understand trends in the data. Some of the vital questions that were answered from the EDA on the dataset include;

1) **What percentage of the Nigerian population has received the first dose of the COVID-19 vaccine?**

As at May 31st 2022; 26,887,802 people in Nigeria have received the first dose of the COVID-19 vaccine. Fig 2 below shows the percentage distribution of the first dose by state populations:

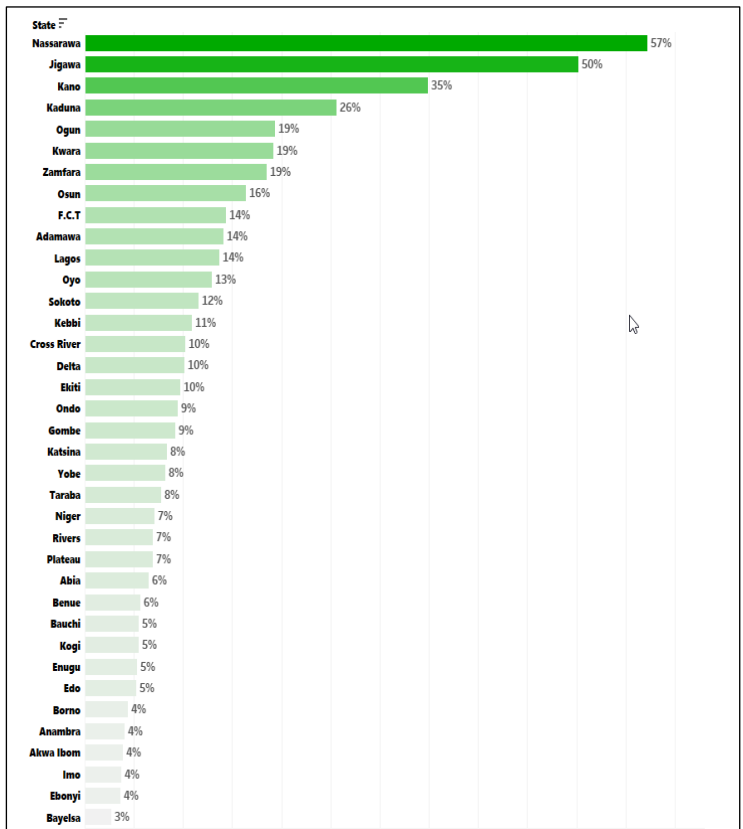

Fig 2: Percentage distribution of the first dose of COVID-19 vaccine by state population

To get this = (First Dose (for each state) * 100) / Population (for each state). From the data, Nassarawa, Jigawa, Kano and Kaduna state are the top 4 states, while Akwa Ibom, Imo, Ebonyi and Bayelsa state are the bottom 4 states for administering the first dose of the vaccine.

2) **How many COVID-19 vaccine doses have been administered in total?**

As at May 31st 2022; the total number of vaccines administered is 45,605,990. See the breakdown by states as shown in fig 3 below;

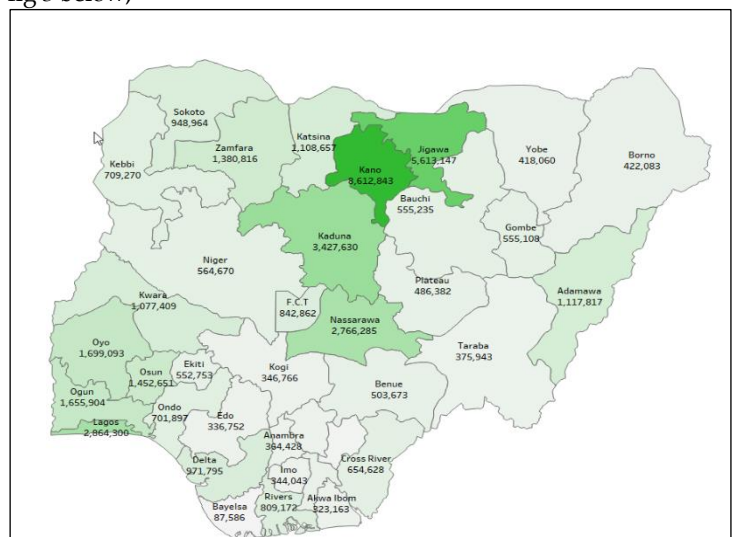

Fig 3: Total number of vaccines administered by states in Nigeria

The total vaccination comprises the total number of people that have received both the first and the second dose of the COVID-19 vaccine. As at May 31st 2022; the data shows that Kano, Jigawa, Kaduna, Lagos, and Nassarawa state led the pack, while Edo, Akwa Ibom, Ebonyi and Bayelsa are at the bottom of the list.

### 3) What percentage of the population has been fully vaccinated?

As at May 31st 2022; only about 10% of Nigerians have been fully vaccinated (first & second dose), and even though this number is relatively small, it is in line with the WHO's SAGE recommendation, which proposed a 3-stage prioritization roadmap for vaccine administration where 1-10 % of a country's total population is expected to be vaccinated at Stage 1. As of May 31st, 2022; Nigeria was still at this stage. A breakdown of the total percentage of vaccinated population by states is shown in Fig 4 below:

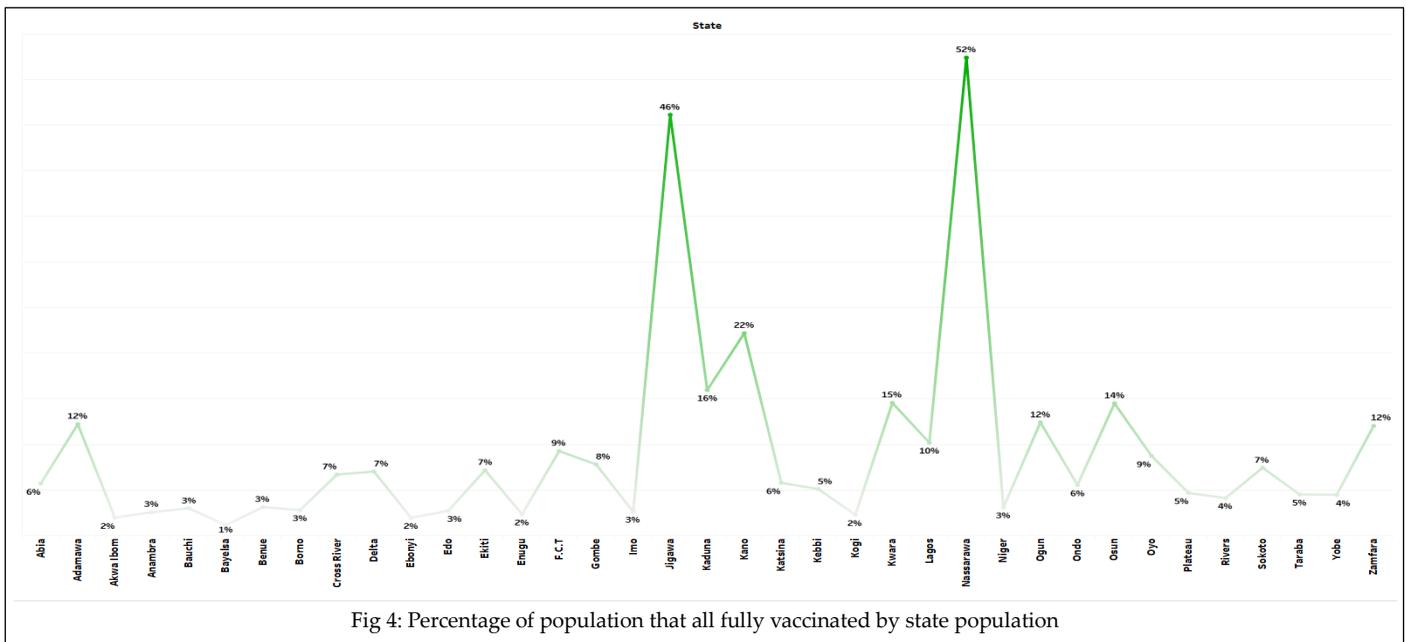

Fig 4: Percentage of population that all fully vaccinated by state population

To get this = (Total Vaccinated Population (for 2nd dose) * 100) / Population (for each state)

Here, Jigawa, Nassarawa, Kano, and Kaduna state lead the pack while Enugu, Kogi, Ebonyi and Bayelsa state are at the bottom.

### 4) What is the percentage difference in the population of fully vaccinated vs partially vaccinated?

As at May 31st 2022; the difference in the population of fully vaccinated (first & second dose) vs partially vaccinated (first dose only) is 8,169,614. This number entails 18% of the population have taken the first dose but not received the second dose. See the breakdown by states as shown in fig 5 below;

| State | % of 1st dose | % of 2nd dose | Difference |
|---|---|---|---|
| Kano | 35% | 22% | 13% |
| Kaduna | 26% | 16% | 10% |
| Ogun | 19% | 12% | 7% |
| Zamfara | 19% | 12% | 7% |
| Kebbi | 11% | 5% | 6% |
| F.C.T | 14% | 9% | 5% |
| Nassarawa | 57% | 52% | 5% |
| Kwara | 19% | 15% | 5% |
| Oyo | 13% | 9% | 4% |
| Sokoto | 12% | 7% | 4% |
| Jigawa | 50% | 46% | 4% |
| Niger | 7% | 3% | 4% |
| Ondo | 9% | 6% | 4% |
| Yobe | 8% | 4% | 4% |
| Cross River | 10% | 7% | 4% |
| Lagos | 14% | 10% | 4% |
| Taraba | 8% | 5% | 3% |
| Kogi | 5% | 2% | 3% |
| Delta | 10% | 7% | 3% |
| Enugu | 5% | 2% | 3% |
| Rivers | 7% | 4% | 3% |
| Katsina | 8% | 6% | 3% |
| Ekiti | 10% | 7% | 3% |
| Edo | 5% | 3% | 3% |
| Benue | 6% | 3% | 3% |
| Bauchi | 5% | 3% | 3% |
| Plateau | 7% | 5% | 2% |
| Akwa Ibom | 4% | 2% | 2% |
| Adamawa | 14% | 12% | 2% |
| Osun | 16% | 14% | 2% |
| Ebonyi | 4% | 2% | 2% |
| Borno | 4% | 3% | 2% |
| Bayelsa | 3% | 1% | 2% |
| Anambra | 4% | 3% | 1% |
| Gombe | 9% | 8% | 1% |
| Imo | 4% | 3% | 1% |
| Abia | 6% | 6% | 1% |

Fig 5: Percentage difference between first and second dose by states

To get this = (1st dose - 2nd dose)*100 / Total Vaccinated Population (for each state)

Here, the data show that Kano, Kaduna, Ogun and Zamfara state have the highest percentage difference between fully vaccinated and partially vaccinated while Anambra, Gombe, Imo and Abia state have the least difference (this can also be traced to their limited number of vaccines distributed so far).

5) Dashboard

Our dashboard contains all the details of the analysis as shown in fig 6 below:

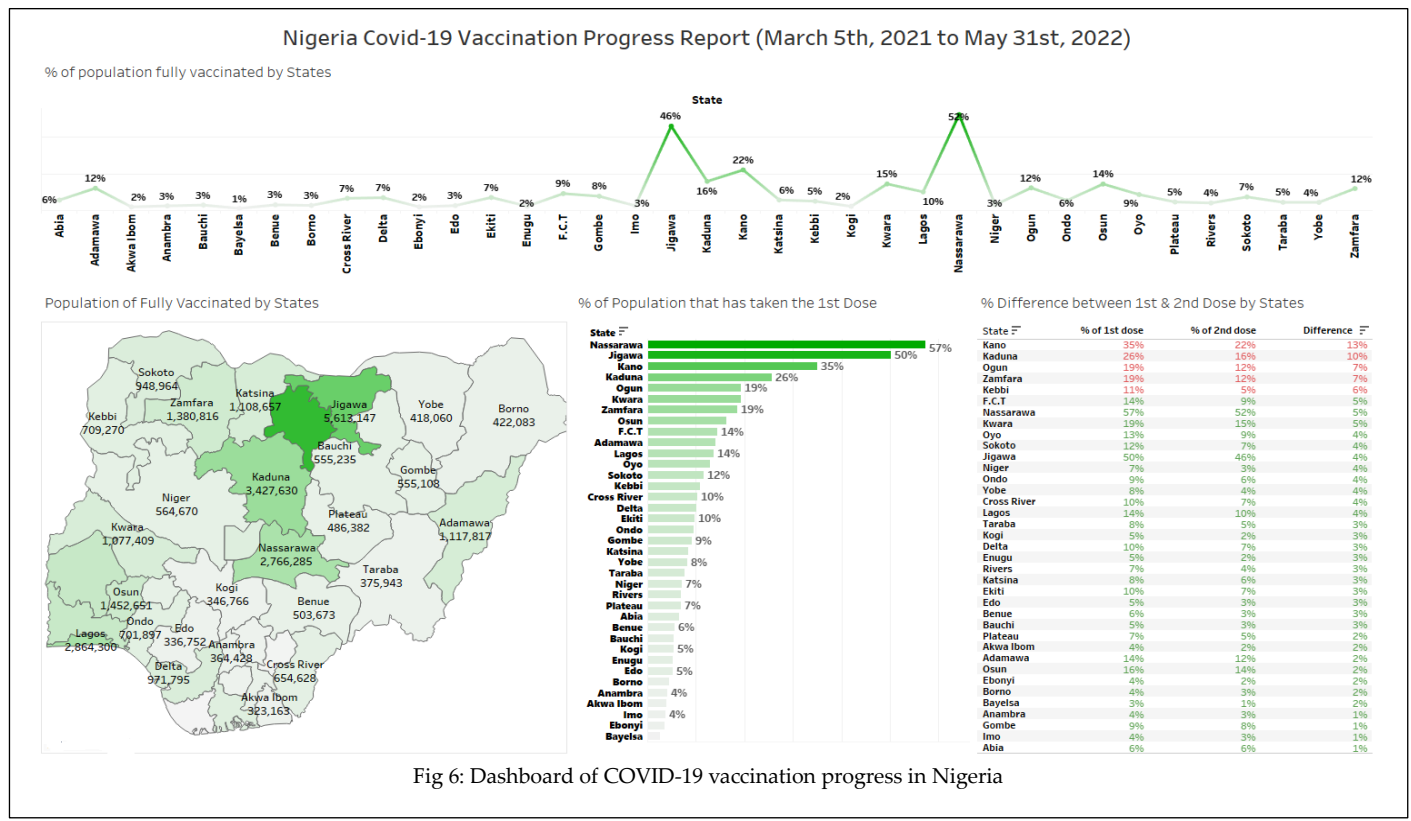

Fig 6: Dashboard of COVID-19 vaccination progress in Nigeria

Finally, the results from this analysis is promising and it also goes to encourage the government agencies to continue to improve their processes to vaccinate more people. However, it took the country 15 months (March, 2021-May, 2022) to finish the stage 1 (i.e 1-10% of population) of the WHO's prioritization roadmap for COVID-19 vaccine distribution. As the country enters the stage 2 of the WHO roadmap to vaccinate 11-20% more of the population, we recommend that the various healthcare institutions incharge of the process become more visible by taking the vaccine to the local communities via the primary healthcare centers and also make more effort to sensitize the people especially in states like Bayelsa, Ebonyi, Imo, Akwa Ibom and Anambra that has only vaccinated less than 5% of their population. We also recommend that government should encourage those that are partially vaccinated (first dose only) to quickly take their second dose to gain more immunity over the COVID-19 disease.

## 5 CONCLUSION

Data is now at the center of most academic and industrial research; it is also the bedrock of modern technological breakthroughs. The COVID-19 pandemic has sparked an enormous interest in data analytics and visualizations. People want to stay informed and understand how the cases are progressing and the impact the virus might have on themselves and their community. Data scientists and analysts from a broad range of domains and skillsets are doing impactful work through exploratory data analysis (EDA) to gain insights, find patterns and trends in data. This paper presented an EDA of the COVID-19 vaccination progress in Nigeria as reported between March 5th, 2021, and May 31st, 2022. We used the R tidyverse package to carry out the data cleaning, preprocessing, and manipulation tasks while Tableau was used to create the visuals and dashboard. This project showcased the effectiveness of data analytics tools like R and Tableau in processing data and presenting insights through visuals/dashboards. These tools can be very effective when dealing with vast amounts of data, especially unstructured data. Most of the tools can be used by non-programmers, thus making it very useful for researchers in all fields. This research has also contributed to the growing literature on COVID-19 research by highlighting the current trends and distribution of the COVID-19 vaccines according to each state in Nigeria. For future work, more research should be done to determine issues that hinder the vaccination progress in southern states of Nigeria where majority have only vaccinated less than 5% of their population.

## Additional Information

The datasets analyzed and complete documentation of the Exploratory Data Analysis process is available at: https://bit.ly/3CPdXZL
The dashboard and other visualization can be found here: https://tabsoft.co/3mfVIVU